\documentclass[12pt]{article}

\usepackage{graphicx}
\usepackage{float}
\usepackage{cite}
\usepackage{amsfonts}
\usepackage{amssymb}
\usepackage{amsmath}
\usepackage{relsize}
\usepackage[compatibility=false]{caption}
\usepackage{subcaption}
\usepackage{xcolor}

\def\gtwid{\mathrel{\raise.3ex\hbox{$>$\kern-.75em\lower1ex\hbox{$\sim$}}}}
\def\ltwid{\mathrel{\raise.3ex\hbox{$<$\kern-.75em\lower1ex\hbox{$\sim$}}}}
\def\square{\kern1pt\vbox{\hrule height 1.2pt\hbox{\vrule width 1.2pt\hskip 3pt
   \vbox{\vskip 6pt}\hskip 3pt\vrule width 0.6pt}\hrule height 0.6pt}\kern1pt}

\begin{document}

\begin{titlepage}

\begin{flushright}
CCTP-2026-02 \\
ITCP/2026/02 \\
UFIFT-QG-26-01
\end{flushright}

\vskip 0.5cm

\begin{center}
{\bf Sub-Leading Logarithms for Scalar Potential Models on de Sitter}
\end{center}

\vskip 0.25cm

\begin{center}
S. P. Miao$^{1*}$, N. C. Tsamis$^{2\dagger}$ and R. P. Woodard$^{3\ddagger}$
\end{center}

\begin{center}
\it{$^{1}$ Department of Physics, National Cheng Kung University, \\
No. 1 University Road, Tainan City 70101, TAIWAN}
\end{center}

\begin{center}
\it{$^{2}$ Institute of Theoretical Physics \& Computational Physics, \\
Department of Physics, University of Crete, \\
GR-700 13 Heraklion, HELLAS}
\end{center}

\begin{center}
\it{$^{3}$ Department of Physics, University of Florida,\\
Gainesville, FL 32611, UNITED STATES}
\end{center}

\vspace{0.25cm}

\begin{center}
ABSTRACT
\end{center}
The continual production of long wavelength scalars and gravitons
during inflation injects secular growth into loop corrections which
would be constant in flat space. One typically finds that each additional
factor of the loop counting parameter can induce up to a certain number
of logarithms of the scale factor. Loop corrections that attain this 
number are known as ``leading logarithms''; those with fewer are 
sub-leading. Starobinsky's stochastic formalism has long been known to
reproduce the leading logarithms of scalar potential models. We show 
that the first sub-leading logarithm is captured by applying the 
stochastic formalism to a certain part of the 1-loop effective
potential. This is checked at 2-loops for a massless, minimally coupled 
scalar with a quartic self-interaction on de Sitter background. 

\begin{flushleft}
\end{flushleft}

\vspace{0.5cm}

\begin{flushleft}
$^{*}$ e-mail: spmiao5@mail.ncku.edu.tw \\
$^{\dagger}$ e-mail: tsamis@physics.uoc.gr \\
$^{\ddagger}$ e-mail: woodard@phys.ufl.edu
\end{flushleft}

\end{titlepage}

\section{Introduction}

Quantum loop effects can be understood as the action of classical 
field theory on the source of virtual particles that emerge from 
the vacuum and persist for a brief period of time according to the 
energy-time uncertainty principle. In flat space this process is
independent of time, although it can engender space dependence,
such as the running of the electrodynamic coupling constant, from 
the fact that long wavelength particles persist longer than those
of short wavelength. The situation is far different in the geometry
of cosmology,
\begin{equation}
ds^2 = -dt^2 \!+\! a^2(t) d\vec{x} \!\cdot\! d\vec{x} = a^2 [-d\eta^2
\!+\! d\vec{x} \!\cdot\! d\vec{x}] \quad \Longrightarrow \quad H(t)
\equiv \tfrac{\dot{a}}{a} \; , \; \epsilon(t) \equiv -
\tfrac{\dot{H}}{H^2} \; . \label{geometry}
\end{equation}
The cosmological expansion continually redshifts ultraviolet momenta
to the infrared, which lengthens the persistence time. For the case
of accelerated expansion ($\epsilon < 1$) a sufficiently long wavelength
particle of zero mass can persist forever. Massless particles that are 
not conformally invariant experience explosive particle production 
during primordial inflation. For example, during de Sitter ($\epsilon = 
0$) the occupation number of a scalar with co-moving wave vector 
$\vec{k}$ grows like \cite{Woodard:2014jba},
\begin{equation}
N(t,\vec{k}) = \Bigl[\tfrac{H a(t)}{2 k}\Bigr]^2 \; . \label{occupation}
\end{equation}
The same result applies for each polarization of gravitons. This is
what caused the primordial power spectra of gravitons 
\cite{Starobinsky:1979ty} and scalars \cite{Mukhanov:1981xt}.

The continual redshift of ultraviolet quanta to the infrared 
during inflation causes the correlators of interacting scalars and
gravitons to become time dependent \cite{Ford:1984hs,Anderson:2000wx,
Weinberg:2005vy,Weinberg:2006ac,Seery:2010kh,Burgess:2010dd,
Kahya:2010xh,Garbrecht:2013coa,Garbrecht:2014dca,Onemli:2015pma,
Karakaya:2017evp,Akhmedov:2013vka}. For example, if a massless, 
minimally coupled scalar is given a self-interaction $V(\phi) = 
\frac1{4!} \lambda \phi^4$ on de Sitter background, and then released 
at $t=0$ in a perturbatively corrected Bunch-Davies vacuum, the 
expectation value of its stress tensor takes the perfect fluid form 
with energy density and pressure \cite{Kahya:2009sz},
\begin{eqnarray}
\rho(t) &\!\!\! = \!\!\!& \tfrac{\lambda H^4}{(4\pi)^4} \Bigl\{2 
\ln^2[a(t)] + \tfrac{13}{6} \ln[a(t)] - \tfrac{43}{18} + \tfrac{\pi^2}{3}
\Bigr\} + O(\lambda^2) \; , \label{rho} \qquad \\
p(t) &\!\!\! = \!\!\!& \tfrac{\lambda H^4}{(4\pi)^4} \Bigl\{-2 
\ln^2[a(t)] - \tfrac{7}{2} \ln[a(t)] + \tfrac{5}{3} - \tfrac{\pi^2}{3}
\Bigr\} + O(\lambda^2) \; . \label{pressure} \qquad 
\end{eqnarray}
It turns out that each additional power of the coupling constant in
this model can produce up to two factors of $\ln[a(t)]$ 
\cite{Tsamis:2005hd}. Corrections that saturate this limit are known 
as {\it leading logarithm}, while those with fewer factors of $\ln[a(t)]$ 
are {\it sub-leading}. For example, the factors of $\ln^2[a(t)]$ in 
expressions (\ref{rho}-\ref{pressure}) are leading logarithm and the 
factors of $\ln[a(t)]$ represent the first sub-leading logarithm. The 
leading logarithm corrections at order $\lambda^2$ would be $\ln^4[a(t)]$, 
and the first sub-leading logarithms would be $\ln^3[a(t)]$.

The counting for quantum gravity is that each additional power of the
inflationary loop-counting parameter $G H^2$ can produce up to a single 
factor of $\ln[a(t)]$. During a long period of inflation these leading
logarithm corrections must grow nonperturbatively strong. This observation
motivates the proposal \cite{Tsamis:1996qq,Tsamis:2011ep} that there is
no scalar inflaton, with inflation instead triggered by a large, positive
cosmological constant. This leads to a long period of nearly de Sitter 
inflation that is eventually brought to an end by the gradual buildup of 
gravitational self-interaction between the vast ensemble of cosmological 
gravitons whose production is reflected in expression (\ref{occupation}). 
Slowing inflation is a leading logarithm effect, occurring when $G H^2 
\ln[a(t)] \sim 1$, and it does seem to happen in the nonperturbative, 
leading logarithm equations which have recently been derived 
\cite{Miao:2024shs,Miao:2025gzm,Miao:2025bmd}. However, the leading 
logarithm geometry is homogeneous and isotropic; primordial perturbations 
require the first sub-leading logarithm, which is down by the correct 
factor of $G H^2$. The purpose of this work is to develop a formalism for 
describing the first sub-leading logarithm. To simplify the analysis, we 
will work in the context of scalar potential models rather than quantum 
gravity. 

A stochastic formalism proposed by Starobinsky \cite{Starobinsky:1986fx}
captures the leading logarithms of scalar potential models at each order
in perturbation theory and can even be used to infer the late time limit
of the series of leading logarithms \cite{Starobinsky:1994bd}. One can
derive Starobinsky's formalism by first integrating the dimensionally 
regulated, Heisenberg field equation to reach the Yang-Feldman equation 
\cite{Tsamis:2005hd},
\begin{equation}
\phi(x) = \phi_0(x) - \int \!\! d^Dx' \sqrt{-g(x')} \, i\theta(\Delta x^0)
\Bigl[\phi_0(x),\phi_0(x')\Bigr] V'\Bigl( \phi(x')\Bigr) \; , \label{YangF}
\end{equation}
where $\phi_0(x)$ is the free field. Iterating this equation would produce 
the usual expansion of the full field $\phi(x)$ in powers of the free 
field. The formalism is still exact at this stage. In particular, $\phi(x)$ 
obeys a 2nd order field equation, it is ``quantum'' in the sense of
failing to commute at timelike separations, and its correlators harbor 
ultraviolet divergences. The passage to Starobinsky's formalism is motivated
by two crucial observations:
\begin{enumerate}
\item{Reaching leading logarithm order requires every pair of free fields
--- including those in the vertex integration --- to contribute a factor
of $\ln[a(t)]$; and}
\item{Factors of $\ln[a(t)]$ come entirely from the infrared portion of
the free field mode sum, and from the leading infrared truncation of the
free field mode function.}
\end{enumerate}
These two facts mean that the leading logarithm contributions to correlators
are not changed if we take $D=4$ and replace the free field by its stochastic 
truncation, 
\begin{eqnarray}
\phi_0(t,\vec{x}) &\!\!\! \longrightarrow \!\!\!& \varphi_0(t,\vec{x}) \equiv
\int_{H}^{H a(t)} \!\!\!\! \tfrac{d^3k}{(2\pi)^3} \tfrac{H}{\sqrt{2 k^3}} 
\Bigl[ \alpha(\vec{k}) e^{-i \vec{k} \cdot \vec{x}} + 
\alpha^{\dagger}(\vec{k}) e^{i \vec{k} \cdot \vec{x}} \Bigr] \; , 
\qquad \label{stochmodesum} \\
& & \Bigl[ \phi_0(t,\vec{x}), \phi_0(t',\vec{x}')\Bigr] \longrightarrow
-\tfrac{i \delta^3(\vec{x} - \vec{x}')}{3 H a^3(t')} \; . \qquad
\end{eqnarray}
Making these two substitutions in the Yang-Feldman equation (\ref{YangF})
results in a relation for a stochastic random field $\varphi(t,\vec{x})$
whose correlators agree with those of $\phi(t,\vec{x})$ at leading 
logarithm order,
\begin{equation}
\varphi(t,\vec{x}) = \varphi_0(t,\vec{x}) - \tfrac1{3H} \! \int_{0}^{t} 
\!\! dt' \, V'\Bigl( \varphi(t',\vec{x}) \Bigr) \; . \label{Langevin}
\end{equation}
Despite agreeing with $\phi(t,\vec{x})$ at leading logarithm order,
$\varphi(t,\vec{x})$ differs in every other respect. In particular, it 
obeys only a first order equation, it is ``classical'' in the sense of 
commuting with itself, and its correlators are completely free of 
ultraviolet divergences.

The preceding discussion makes clear that reaching leading logarithm
order requires each pair of free fields $\phi_0$ to contribute an
infrared logarithm. To get the first sub-leading logarithm we need to
``waste'' a single pair of free fields. This suggests that we include
the 1-loop effective potential in the integrated form (\ref{Langevin})
of Starobinsky's Langevin equation. This quantity turns out to require
renormalization, to be time dependent, and to include terms which are 
actually leading logarithm order. However, we show that a certain 
portion of the renormalized 1-loop effective potential indeed captures 
the first sub-leading logarithm corrections. 

This paper has five sections, of which this Introduction is the first.
In Section 2 we give the Feynman rules and work out the two 1-loop 
counterterms that are needed to renormalize the 1-loop effective
potential. That effective potential is computed in Section 3. We also
isolate the part of it that produces sub-leading logarithms and we 
employ it in the integrated form of Starobinsky's Langevin equation 
(\ref{Langevin}) to predict the leading and sub-leading logarithms of 
the expectation value of $\phi^2(x)$ at order $\lambda$. That 
prediction is confirmed in Section 4, up to some ambiguity about the
lower limit of the stochastic mode sum (\ref{stochmodesum}). Our 
conclusions comprise Section 5. 

\section{Counterterms}

The purpose of this section is to work out the counterterms needed to
renormalize the 1-loop correction to the effective potential. We begin 
by reviewing the Feynman rules. Then the 1-loop self-mass is computed 
to determine the conformal counterterm. The section closes with an 
evaluation of the 4-point vertex at 1-loop order to determine the 
coupling constant counterterm.

\subsection{Feynman Rules} 

The Lagrangian for a massless, minimally coupled scalar with a quartic
self-interaction is, 
\begin{eqnarray}
\lefteqn{\mathcal{L} = -\tfrac12 \partial_{\mu} \phi \partial_{\nu} \phi
g^{\mu\nu} \sqrt{-g} -\tfrac1{4!} \lambda \phi^4 \sqrt{-g} } \nonumber \\
& & \hspace{2.5cm} -\tfrac12 \delta Z \partial_{\mu} \phi \partial_{\nu} \phi
g^{\mu\nu} \sqrt{-g} - \tfrac12 \delta \xi \phi^2 R \sqrt{-g} - \tfrac1{4!}
\delta \lambda \phi^4 \sqrt{-g} \; . \qquad \label{Lagrangian}
\end{eqnarray}
The first line is primitive whereas the second line consists of counterterms.
Most of the propagator can be expressed in terms of the de Sitter invariant
length function,
\begin{equation}
y(x;x') \equiv a a' H^2 \Bigl[ \Vert \vec{x} \!-\! \vec{x}' \Vert^2 -
(\vert\eta \!-\! \eta' \vert \!-\! i \epsilon)^2 \Bigr] \; . \label{ydef}
\end{equation}
After the appropriate infrared subtraction the propagator is 
\cite{Janssen:2008px},
\begin{eqnarray}
\lefteqn{i\Delta(x;x') = \tfrac{H^{D-2}}{(4\pi)^{\frac{D}2}} \Bigl\{
\tfrac{\Gamma(\frac{D}2)}{\frac{D}2 - 1} (\tfrac{4}{y})^{\frac{D}2 -1}
+ \tfrac{\Gamma(\frac{D}2 +1)}{\frac{D}2 - 2} (\tfrac{4}{y})^{\frac{D}2 -2}
\Bigr\} + k \Bigl\{ \ln(a a') + \Psi(D)\Bigr\} } \nonumber \\
& & \hspace{2.5cm} + \tfrac{H^{D-2}}{(4\pi)^{\frac{D}2}} \sum_{n=1}^{\infty}
\Bigl\{ \tfrac{\Gamma(D - 1 + n)}{n \Gamma(\frac{D}2 + n)} (\tfrac{y}{4})^n 
- \tfrac{\Gamma(\frac{D}2 + 1 + n)}{(n - \frac{D}2 + 2) (n+1)!}
(\tfrac{y}{4})^{n-\frac{D}2 +2} \Bigr\} \; , \qquad \label{propagator}
\end{eqnarray}
where we define,
\begin{equation}
k \equiv \tfrac{H^{D-2}}{(4\pi)^{\frac{D}2}} \tfrac{\Gamma(D-1)}{
\Gamma(\frac{D}2)} \qquad , \qquad \Psi(D) \equiv -\psi(1 - \tfrac{D}2) + 
\psi(\tfrac{D}2 - \tfrac12) + \psi(D-1) - \gamma \; . \label{constants}
\end{equation}
Note that only the terms on the first line survive in $D=4$ dimensions.

\subsection{The Conformal Counterterm}

\begin{figure}[H]
\centering
\vskip 1cm
\hskip -1cm
\includegraphics[width=10cm]{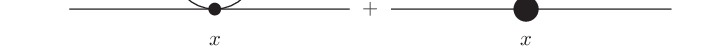}
\caption{\footnotesize 1-loop contributions to the scalar self-mass.
The leftmost diagram is the primitive contribution and the rightmost 
one is the conformal counterterm.}
\label{selfmass}
\end{figure}

The diagrams which comprise the 1-loop self-mass are shown in 
Figure~\ref{selfmass}.
\noindent The corresponding analytic expression is, 
\begin{equation}
-i M^2(x;x') = \tfrac12 \times -i\lambda \sqrt{-g} \, \delta^D(x - x') 
\times i\Delta(x;x) - i \delta \xi R \sqrt{-g} \, \delta^D(x - x') \; .
\label{Msq}
\end{equation}
The dimensionally regulated coincidence limit of the propagator is,
\begin{equation}
i\Delta(x;x) = k [2 \ln(a) + \Psi(D)] \; . \label{coincidence}
\end{equation}
The Ricci scalar is $R = D (D-1) H^2$, and it follows that the 
conformal counterterm is,
\begin{equation}
\delta \xi = -\tfrac{\lambda \mu_1^{D-4}}{(4\pi)^{\frac{D}2}} 
\tfrac{\Gamma(D-1)}{\Gamma(\frac{D}2)} \tfrac{\Psi(D)}{2 D (D-1)} 
+ O(\lambda^2) \; . \label{deltaxi}
\end{equation}
The arbitrary finite part is represented by the regularization scale
$\mu_1$.

\subsection{The Vertex Counterterm}

\begin{figure}[H]
\centering
\vskip 1cm
\hskip -3cm
\includegraphics[width=12cm]{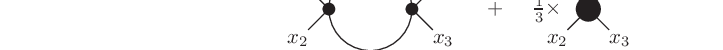}
\caption{\footnotesize 1-loop contributions to the $s$-channel vertex
function. The leftmost diagram is the primitive contribution and the 
rightmost one is the vertex counterterm.}
\label{vertex}
\end{figure}

The 1-loop corrections to the vertex function include $s$-channel,
$t$-channel and $u$-channel contributions. The diagrams that comprise
the $s$ channel are depicted in Figure~\ref{vertex}. 
\noindent The corresponding analytic expression is,
\begin{eqnarray}
\lefteqn{ -i V_{s}(x_1;\dots) = \tfrac12 \!\times\! (-i\lambda)^2 
\sqrt{-g} \, \delta^D(x_1 \!-\! x_2) \!\times\! [i\Delta(x_1;x_3)]^2 \!
\times\! \sqrt{-g} \, \delta^D(x_3 \!-\! x_4) } \nonumber \\
& & \hspace{3.2cm} + \tfrac13 \!\times\! -i \delta \lambda \sqrt{-g} \, 
\delta^D(x_1 \!-\! x_2) \delta^D(x_1 \!-\! x_3) \delta^D(x_1 \!-\! x_4) 
\; . \qquad \label{Vs}
\end{eqnarray}

The ultraviolet divergence depends on the square of the propagator, 
which will be treated generically because it is used as well in
Section 4. This quantity is logarithmically divergent so the 
dimensional regularization only needs to be kept for the most singular 
term,
\begin{eqnarray}
\lefteqn{ [i\Delta(x;x')]^2 = \tfrac{\Gamma^2(\frac{D}2 - 1)}{16 \pi^D}
\tfrac{1}{[a a' \Delta x^2]^{D-2}} - 2 \!\times\! \tfrac{1}{4 \pi^2 a a' 
\Delta x^2} } \nonumber \\
& & \hspace{0cm} \times \tfrac{H^2}{8 \pi^2} \Bigl[\ln(H^2 \Delta x^2) 
\!+\! 2 \gamma \!-\! 2\Bigr] + \tfrac{H^4}{64 \pi^4} \Bigl[
\ln(H^2 \Delta x^2) \!+\! 2 \gamma \!-\! 2\Bigr]^2 + O(D\!-\!4) \; .
\qquad \label{initialexp}
\end{eqnarray}
One isolates the divergence by first extracting a flat space d'Alembertian
and then adding zero in the form of the flat space propagator equation
\cite{Onemli:2002hr},
\begin{eqnarray}
\tfrac{1}{\Delta x^{2D-4}} &\!\!\! = \!\!\!& \tfrac{\partial^2}{2(D-3)(D-4)}
\tfrac{1}{\Delta x^{2D-6}} \; , \qquad \\
&\!\!\! = \!\!\!& \tfrac{\partial^2}{2 (D-3) (D-4)} \Bigl[
\tfrac{1}{\Delta x^{2D-6}} - \tfrac{\mu^{D-4}}{\Delta x^{D-2}} \Bigr] +
\tfrac{\mu^{D-4}}{2(D-3)(D-4)} \tfrac{4 \pi^{\frac{D}2} i \delta^D(x-x')}{
\Gamma(\frac{D}2 - 1)} \; , \qquad \\
&\!\!\! = \!\!\!& \tfrac{\mu^{D-4}}{2(D-3)(D-4)} \tfrac{4 \pi^{\frac{D}2} 
i \delta^D(x-x')}{\Gamma(\frac{D}2 - 1)} - \tfrac14 \partial^2 \Bigl[
\tfrac{\ln(\mu^2 \Delta x^2)}{\Delta x^2}\Bigr] + O(D\!-\!4) \; . \qquad 
\label{extract}
\end{eqnarray}
Substituting (\ref{extract}) in (\ref{initialexp}) gives,
\begin{eqnarray}
\lefteqn{[i\Delta(x;x')]^2 = \tfrac{\mu^{D-4}}{8 \pi^{\frac{D}2}} 
\tfrac{\Gamma(\frac{D}2 -1)}{(D-3) (D-4)} \tfrac{i \delta^D(x - x')}{
(a a')^{D-2}} - \tfrac{1}{64 \pi^4} \tfrac{\partial^2}{(a a')^2} \Bigl[ 
\tfrac{\ln(\mu^2 \Delta x^2)}{\Delta x^2} \Bigr] - \tfrac{H^2}{16 \pi^4 
a a' \Delta x^2} } \nonumber \\
& & \hspace{0cm} \times \Bigl[\ln(H^2 \Delta x^2) \!+\! 2 \gamma \!-\! 2
\Bigr] + \tfrac{H^4}{64 \pi^4} \Bigl[\ln(H^2 \Delta x^2) \!+\! 2 \gamma 
\!-\! 2\Bigr]^2 + O(D\!-\!4) \; . \qquad \label{propsq} 
\end{eqnarray}
It follows that the coupling constant counterterm is,
\begin{equation}
\delta \lambda = -\tfrac{3 \lambda^2 \mu_2^{D-4}}{16 \pi^{\frac{D}2}}
\tfrac{\Gamma(\tfrac{D}2 - 1)}{(D-3) (D-4)} + O(\lambda^3) \; . 
\label{deltalambda}
\end{equation}
Note that we use a mass scale $\mu_2$ which can differ from the scale
$\mu_1$ employed for the conformal counterterm (\ref{deltaxi}). 

\section{$V_{\rm eff}(\phi)$ at 1-Loop}

The purpose of this section is to compute the 1-loop effective potential
and use it in equation (\ref{Langevin}) to predict the first sub-leading
logarithm corrections to correlators of $\phi(x)$. We begin by discussing
the propagator in the presence of a constant scalar background. This
propagator contains a de Sitter breaking, time dependent part because it
must recover expression (\ref{propagator}) in the limit of zero background
field. The next step is to compute the effective potential, removing its 
ultraviolet divergences using the counterterms (\ref{deltaxi}) and 
(\ref{deltalambda}). Because the propagator includes even the infrared 
parts of the mode sum and mode functions, part of this effective potential
includes leading logarithm effects. We subtract these and isolate the 
portions that produce the first sub-leading order. The section closes by
predicting first sub-leading logarithms contributions to the expectation 
value of $\phi^{2}$ at order $\lambda$.

\subsection{The Propagator in a Constant Background}

The 1-loop correction to $V(\phi) = \frac{\lambda}{4!} \phi^4$ is,
\begin{equation}
V'_{\rm eff}(\phi) = \tfrac{\lambda \phi}{2} i \Delta[\phi](x;x) +
\delta \xi R \phi + \tfrac{\delta \lambda}{6} \phi^3 \; , 
\label{Veffinitial}
\end{equation}
where $i\Delta[\phi](x;x')$ is the propagator in the presence of a 
constant scalar background. This background corresponds to a mass of
$m^2 = \frac{\lambda}{2} \phi^2$. A formal expression for this propagator
was first derived by Chernikov and Tagirov in 1968 \cite{Chernikov:1968zm}
in terms of $\mbox{}_{2}F_{1}(\frac{D-1}{2} + \nu,\frac{D-1}{2} -\nu;
\frac{D}2;1-\frac{y}{4})$, where the index $\nu$ is,
\begin{equation}
\nu^2 \equiv (\tfrac{D-1}{2})^2 - \tfrac{\lambda \phi^2}{2 H^2} \; .
\label{nudef}
\end{equation}
However, this expression becomes infrared divergent in the limit of 
vanishing background. This can be eliminated by working on a finite 
spatial volume, which corresponds to the subtraction of a homogeneous 
term \cite{Janssen:2009pb},
\begin{eqnarray}
\lefteqn{ i\Delta[\phi](x;x') = \tfrac{H^{D-2}}{(4\pi)^{\frac{D}2}} 
\Bigl\{ \tfrac{\Gamma(\frac{D}2)}{\frac{D}2 - 1} (\tfrac{4}{y})^{
\frac{D}2 -1} + \tfrac{\Gamma(2\nu) \Gamma(\nu)}{\Gamma(\frac12 + \nu)
\Gamma(\frac{D-1}{2})} \tfrac{(a a')^{\nu - (\frac{D-1}2)}}{\nu - 
(\frac{D-1}{2})} + \tfrac{\Gamma(\frac{D}2 -1) \Gamma(2 - \frac{D}2)
}{\Gamma(\frac12 + \nu) \Gamma(\frac12 - \nu)} } \nonumber \\
& & \hspace{0.5cm} \times \sum_{n=0}^{\infty} \Bigl[
\tfrac{\Gamma(\frac32 + \nu + n) \Gamma(\frac32 - \nu + n)}{
\Gamma(3 - \frac{D}2 + n) (n+1)!} (\tfrac{y}{4})^{n-\frac{D}2 +2} -
\tfrac{\Gamma(\frac{D-1}{2} + \nu + n) \Gamma(\frac{D-1}{2} - \nu + n)}{
\Gamma(\frac{D}2 + n) n!} (\tfrac{y}{4})^n \Bigr] \Bigr\} \; . \qquad
\label{propphi}
\end{eqnarray}
The problem of obtaining a smooth small mass limit for this theory 
has also been studied extensively by Kamenshchik and Petriakova 
\cite{Kamenshchik:2024ybm,Kamenshchik:2025ses}. 

\subsection{1-Loop Effective Potential}

In dimensional regularization the coincidence limit of the propagator
(\ref{propphi}) derives from the two $y^0$ terms,
\begin{equation}
i\Delta[\phi](x;x) = \tfrac{H^{D-2}}{(4\pi)^{\frac{D}2}} \Bigl\{
\tfrac{\Gamma(2\nu) \Gamma(\nu)}{\Gamma(\frac12 + \nu)
\Gamma(\frac{D-1}{2})} \tfrac{a^{2\nu - D +1}}{\nu - (\frac{D-1}{2})} 
- \tfrac{\Gamma(\frac{D}2 -1) \Gamma(2 - \frac{D}2)}{\Gamma(\frac{D}2)} 
\tfrac{\Gamma(\frac{D-1}{2} + \nu) \Gamma(\frac{D-1}{2} - \nu )}{
\Gamma(\frac12 + \nu) \Gamma(\frac12 - \nu)} \Bigr\} \; . \label{coinc}
\end{equation}
The key expansion is,
\begin{eqnarray}
\lefteqn{\tfrac{\Gamma(\frac{D-1}{2} + \nu) \Gamma(\frac{D-1}{2} - \nu )}{
\Gamma(\frac12 + \nu) \Gamma(\frac12 - \nu)} = \Bigl[ (\tfrac{D-3}{2})^2
- \nu^2\Bigr] } \nonumber \\
& & \hspace{3cm} \times \Bigl\{1 \!+\! \Bigl[ \psi(\tfrac12 \!+\! \nu) \!+\! 
\psi(\tfrac12 \!-\! \nu)\Bigr] (\tfrac{D-4}{2}) + O\Bigl( (D\!-\!4)^2\Bigr) 
\Bigr\} . \qquad \label{expansion}
\end{eqnarray}
It follows that the coincident propagator is,
\begin{eqnarray}
\lefteqn{ i\Delta[\phi](x;x) = \tfrac{H^{D-2}}{(4\pi)^{\frac{D}2}}
\tfrac{4 \Gamma(3 - \frac{D}2)}{(D-2)(D-4)} \Bigl[-(D\!-\!2) + 
\tfrac{\lambda \phi^2}{2 H^2}\Bigr] + \tfrac{H^2}{16 \pi^2} 
\Bigl\{\tfrac{\Gamma(2\nu) \Gamma(\nu)}{\Gamma(\frac12 + \nu) 
\Gamma(\frac32)} \tfrac{a^{2\nu -3}}{\nu - \frac32} } \nonumber \\
& & \hspace{3cm} + \Bigl[-2 + \tfrac{\lambda \phi^2}{2 H^2}\Bigr] \Bigl[ 
\psi(\tfrac12 + \nu) + \psi(\tfrac12 - \nu)\Bigr] \Bigr\} + O(D\!-\!4) 
\; , \qquad \\
& & = \tfrac{H^{D-2}}{(4\pi)^{\frac{D}2}}
\tfrac{4 \Gamma(3 - \frac{D}2)}{(D-2)(D-4)} \Bigl[-(D\!-\!2) + 
\tfrac{\lambda \phi^2}{2 H^2}\Bigr] + \tfrac{H^2}{16 \pi^2} 
\Bigl\{2 \Bigl[\tfrac{\Gamma(\nu)}{\Gamma(\frac32)}\Bigr]^2 
\tfrac{(2 a)^{2\nu -3}}{\nu - \frac32} - \tfrac{2}{\nu - \frac32}
\nonumber \\
& & \hspace{2cm} -2 \nu \!-\! 2 \!+\! \Bigl[-2 + \tfrac{\lambda \phi^2}{
2 H^2} \Bigr] \Bigl[ \psi(\tfrac12 \!+\! \nu) \!+\! \psi(\tfrac52 \!-\! 
\nu)\Bigr] \Bigr\} \!+\! O(D\!-\!4) \; . \qquad \label{coincexp}
\end{eqnarray}

The next step is to substitute (\ref{coincexp}) in expression
(\ref{Veffinitial}) for the effective potential. The potentially
divergent parts are,
\begin{eqnarray}
\lefteqn{ \tfrac{\lambda \phi H^{D-2}}{(4\pi)^{\frac{D}2}} 
\tfrac{\Gamma(3 - \frac{D}2)}{(D-2) (D-4)} \Bigl[-2 (D\!-\!2) \!+\!
\tfrac{\lambda \phi^2}{H^2} \Bigr] - \tfrac{\lambda \phi H^2 
\mu_1^{D-4}}{(4\pi)^{\frac{D}2}} \tfrac{\Psi(D) \Gamma(D-1)}{2
\Gamma(\frac{D}2)} } \nonumber \\
& & \hspace{8cm} - \tfrac{\lambda^2 \phi^3 \mu_2^{D-4}}{32 
\pi^{\frac{D}2}} \tfrac{\Gamma(\frac{D}2 -1)}{(D-3) (D-4)} \; , 
\qquad \\
& & \hspace{1.5cm} = \tfrac{\lambda \phi H^2}{16 \pi^2} \Bigl\{2
\ln(\tfrac{2 \mu_1}{H}) \!-\! \tfrac12\Bigr\} - 
\tfrac{\lambda^2 \phi^3}{32 \pi^2} \Bigl\{ \ln(\tfrac{2 \mu_2}{H})
\!-\! \tfrac12 \!-\! \gamma\Bigr\} + O(D\!-\!4) \; . \qquad \label{Veffdiv}
\end{eqnarray}
It follows that the 1-loop effective potential is ,
\begin{eqnarray}
\lefteqn{V'_{\rm eff} = \tfrac{\lambda H^2 \phi}{16 \pi^2} \Bigl\{ 
2 \ln(\tfrac{2 \mu_1}{H}) \!-\! \tfrac32 \!-\! \nu \!-\!
\psi(\tfrac12 \!+\! \nu) \!-\! \psi(\tfrac52 \!-\! \nu) \!+\! \Bigl[
\tfrac{\Gamma(\nu)}{\Gamma(\frac32)}\Bigr]^2 \tfrac{(2 a)^{2\nu - 3}}{
\nu - \frac32} \!-\! \tfrac{1}{\nu - \frac32} \Bigr\} } \nonumber \\
& & \hspace{3.8cm} - \tfrac{\lambda^2 \phi^3}{64 \pi^2} \Bigl\{ 2 
\ln(\tfrac{2 \mu_2}{H}) \!-\! 1 \!-\! 2 \gamma \!-\! \psi(\tfrac12 
\!+\! \nu) \!-\! \psi(\tfrac52 \!-\! \nu)\Bigr\} . \qquad \label{Veff}
\end{eqnarray}
Aside from some different renormalization conventions, this is the same
result as the $\epsilon \rightarrow 0$ limit of the ``Hubble effective 
potential'' derived in 2009 \cite{Janssen:2009pb}.

\subsection{First Sub-Leading Logarithm Contribution}

In $D=4$ the index can be expanded in powers of the scalar,
\begin{equation}
\nu = \tfrac32 \sqrt{1 - \tfrac{2 \lambda \phi^2}{9 H^2}} = \tfrac32
- \tfrac{\lambda \phi^2}{6 H^2} \Bigl\{1 + \tfrac12 \tfrac{\lambda 
\phi^2}{9 H^2} + \tfrac12 (\tfrac{\lambda \phi^2}{9 H^2})^2 + 
\tfrac58 (\tfrac{\lambda \phi^2}{9 H^2})^3 + \dots \Bigr\} \; .
\label{nuexp}
\end{equation}
We can also expand the time-dependent part of (\ref{Veff}),
\begin{eqnarray}
\lefteqn{\Bigl[\tfrac{\Gamma(\nu)}{\Gamma(\frac32)}\Bigr]^2 
\tfrac{(2 a)^{2\nu - 3}}{\nu - \frac32} \!-\! \tfrac{1}{\nu - \frac32} =
\tfrac1{\nu - \frac32} \Bigl( \Bigl[ \tfrac{\Gamma(\nu)}{\Gamma(\frac32)}
\Bigr]^2 - 1\Bigr) + \Bigl[ \tfrac{\Gamma(\nu)}{\Gamma(\frac32)}\Bigr]^2
\tfrac{[(2 a)^{2\nu - 3} - 1]}{\nu - \frac32} \; , } \\
& & \hspace{2.8cm} = \tfrac{[\Gamma(\nu)/\Gamma(\frac32)]^2 - 1}{
\nu - \frac32} + 2 \ln(2 a) \Bigl[ \tfrac{\Gamma(\nu)}{\Gamma(\frac32)}
\Bigr]^2 \sum_{n=1}^{\infty} \tfrac{[(2 \nu - 3) \ln(2 a)]^{n-1}}{n!} 
\; . \qquad \label{timexp}
\end{eqnarray}
Expression (\ref{Veff}) can be rewritten as,
\begin{eqnarray}
\lefteqn{V'_{\rm eff}(\phi) = \tfrac{\lambda H^2 \phi}{16 \pi^2} 
\Bigl\{ 2 \ln(\tfrac{2 \mu_1}{H}) \!-\! \tfrac32 \!-\! \nu  
\!-\! \psi(\tfrac12 \!+\! \nu) \!-\! \psi(\tfrac52 \!-\! \nu) \!+\! 
\tfrac{[\Gamma(\nu)/\Gamma(\frac32)]^2 - 1}{\nu - \frac32} \Bigr\} } 
\nonumber \\
& & \hspace{3cm} - \tfrac{\lambda^2 \phi^3}{64 \pi^2} \Bigl\{ 2 
\ln(\tfrac{2 \mu_2}{H}) \!-\! 1 \!-\! 2 \gamma \!-\! \psi(\tfrac12 
\!+\! \nu) \!-\! \psi(\tfrac52 \!-\! \nu)\Bigr\} \nonumber \\
& & \hspace{5.5cm} + \tfrac{\lambda H^2 \phi \ln(2 a)}{8 \pi^2} \Bigl[ 
\tfrac{\Gamma(\nu)}{\Gamma(\frac32)} \Bigr]^2 \sum_{n=0}^{\infty} 
\tfrac{[(2 \nu - 3) \ln(2 a)]^{n}}{(n+1)!} \; . \qquad \label{complete}
\end{eqnarray}

Employing these expansions in (\ref{Veff}), we see that the 
derivative of the effective potential (\ref{complete}) takes the
general form,
\begin{equation}
V'_{\rm eff}(\phi) = \lambda H^2 \phi \!\times\! 
f(\tfrac{\lambda \phi^2}{H^2}) + \lambda H^2 \phi \ln(2a) \!\times\!
g (\tfrac{\lambda \phi^2}{H^2}) \!\times\! h(\tfrac{\lambda \phi^2 
\ln(2a)}{H^2}) \; . \label{generalform}
\end{equation}
Now recall that it would be double counting to include the effective 
potential in the Feynman rules as a new, fundamental interaction. We 
use it here only in the integrated form of Starobinsky's Langevin 
equation (\ref{Langevin}). When this is iterated, all the fields
eventually become the infrared truncated, free field $\varphi_0(x)$.
Two powers of $\varphi_0$ can produce a factor of $\ln(a)$, and remaining 
at leading logarithm order requires {\it two} factors of $\ln(a)$ for 
each power of $\lambda$. Hence functions of $\lambda \phi^2 \ln(2a)$ 
do not change leading logarithm status, but the insertion of even one
power of $\lambda \phi^2$, without the $\ln(2 a)$ does. It follows 
that we can identify the leading logarithm and first sub-leading 
logarithm parts of $V'_{\rm eff}$,
\begin{eqnarray}
{\rm Leading} &\!\!\! \Longrightarrow \!\!\!& \lambda H^2 \phi 
\ln(2 a) \!\times\! g_0 \!\times\! h(\tfrac{\lambda \phi^2 \ln(2 a)}{
H^2}) \; , \qquad \label{genLeading} \\
{\rm 1st\ Sub-Leading} &\!\!\! \Longrightarrow \!\!\!& \lambda H^2 
\phi \!\times\! \Bigl\{ f_0 + g_1 \!\times\! (\tfrac{\lambda \phi^2 
\ln(2 a)}{H^2}) \!\times\! h(\tfrac{\lambda \phi^2 \ln(2 a)}{H^2}) 
\Bigr\} \; , \qquad \label{genSub}
\end{eqnarray}
where $f_0 \equiv f(0)$, $g_0 \equiv g(0)$ and $g_1 \equiv g'(0)$.
Comparison of expressions (\ref{complete}), (\ref{generalform}) and
(\ref{genLeading}) imply that the leading logarithm contribution is,
\begin{equation}
V'_{{\rm eff},0}(\phi) = \tfrac{\lambda H^2 \phi \ln(2 a)}{8 \pi^2} 
\sum_{n=0}^{\infty} \tfrac1{(n+1)!} \Bigl[ -
\tfrac{\lambda \phi^2 \ln(2a)}{3 H^2} \Bigr]^n \; . \label{Leading}
\end{equation}
Comparing (\ref{complete}), (\ref{generalform}) and (\ref{genSub}) 
gives the first sub-leading contribution,
\begin{equation}
V'_{{\rm eff},1}(\phi) = \tfrac{\lambda H^2 \phi}{16 \pi^2} \Bigl\{
2 \ln(\tfrac{\mu_1}{2 H}) \!+\! \sum_{n=0}^{\infty}
\tfrac{[2 \psi(\frac32) - \frac{n}{3}]}{(n+1)!} \Bigl[ -
\tfrac{\lambda \phi^2 \ln(2a)}{3 H^2} \Bigr]^{n+1} \Bigr\} \; . 
\label{1stSub}
\end{equation}

\begin{figure}[H]
\centering
\vskip 0.5cm
\hskip -3cm
\includegraphics[width=10cm]{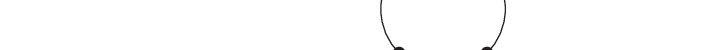}
\vskip .3cm
\caption{\footnotesize One term in the sum over 1PI vacuum graphs 
that contributes to the 1-loop effective potential. Each vertex 
contributes a factor of $\frac{\lambda}2 \phi^2$.}
\label{Vsum}
\end{figure}

It is striking that the 1-loop effective potential includes leading 
logarithm corrections, as well as every order of sub-leading 
correction. The reason for this becomes apparent if one considers the 
effective potential, as Coleman and Weinberg originally did 
\cite{Coleman:1973jx}, as the sum of vacuum 1PI (one-particle-irreducible) 
diagrams like the one depicted in Figure~\ref{Vsum}. 
\noindent The $N$-point contribution contains $N$ factors of 
$\frac{\lambda}{2} \phi^2$ and $N$ propagators. Leading logarithm
contributions arise when all $N$ propagators give a factor of $H^2 
\ln(a)$. The first sub-leading contribution comes from when one of
the propagators fails to produce this factor, the second sub-leading 
contribution comes from two propagators failing to produce this 
factor, and so on.

\subsection{Stochastic Predictions at Order $\lambda$}

Our proposal is to add $V'_{\rm eff,1}$ to $V'$ in the integrated
version of Starobinsky's Langevin equation (\ref{Langevin}),
\begin{equation}
\varphi = \varphi_0 - \tfrac1{3 H} I\Bigl[V'(\varphi) + 
V'_{\rm eff,1}(\varphi) \Bigr] \; . \label{newLang}
\end{equation}
Here we define $I$ acting on a function of time $f(t)$ as the definite
integral,
\begin{equation}
I[f](t) \equiv \int_{0}^{t} \!\! dt' f(t') \; . \label{fdef}
\end{equation}
The solution to order $\lambda$ is,
\begin{equation}
\varphi = \varphi_0 - \tfrac{\lambda}{18 H} I[\varphi_0^3] - 
\tfrac{\lambda H \ln(\frac{\mu_1}{2H})}{24 \pi^2} I[\varphi_0] + 
O(\lambda) \; . \label{solution}
\end{equation}
The square of (\ref{solution}) is,
\begin{equation}
\varphi^2 = \varphi_0^2 - \tfrac{\lambda \varphi_0}{9 H} I[\varphi_0^3]
- \tfrac{\lambda H \ln(\frac{\mu_1}{2H}) \varphi_0}{12 \pi^2} 
I[\varphi_0] + O(\lambda^2) \; . \label{square}
\end{equation}
Taking the expectation value gives,
\begin{equation}
\langle \varphi^2 \rangle = \tfrac{H^2}{4 \pi^2} \ln(a) - 
\tfrac{\lambda H^2}{144 \pi^4} \ln^3(a) - \tfrac{\lambda H^2 
\ln(\frac{\mu_1}{2 H})}{96 \pi^4} \ln^2(a) + O(\lambda^2) \; . 
\label{VEVsquare}
\end{equation} 

\section{$\langle \Omega \vert \phi^2(t,\vec{x}) \vert \Omega \rangle$ at Order $\lambda$}

The purpose of this section is to confirm the stochastic prediction 
for the first sub-leading logarithm contribution to the expectation
value of $\phi^2$. We begin by explaining the Schwinger-Keldysh formalism.
The primitive contribution is then evaluated using dimensional 
regularization. Although removing all divergences from the expectation 
value of $\phi^2$ requires composite operator renormalization, both the
leading logarithm contribution at order $\lambda$ and the first sub-leading
logarithm contribution become ultraviolet finite when the conformal
counterterm (\ref{deltaxi}) is included.

\subsection{The Schwinger-Keldysh Formalism}

\begin{figure}[H]
\centering
\vskip 1cm
\hskip -3cm
\includegraphics[width=10cm]{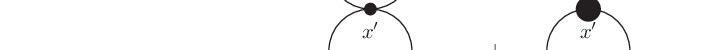}
\vskip .8cm
\caption{\footnotesize Order $\lambda$ contributions to $\langle \Omega \vert
\phi^2(x) \vert \Omega \rangle$. The leftmost diagram is the primitive 
contribution and the rightmost one is from the conformal counterterm.}
\label{phisq}
\end{figure}

The order $\lambda$ contribution to the expectation value of $\phi^2$
was first computed in 1982 \cite{Vilenkin:1982wt,Linde:1982uu,
Starobinsky:1982ee}, and agrees with the stochastic prediction
(\ref{VEVsquare}). The order $\lambda$ contributions are shown in 
Figure~\ref{phisq}.
\noindent Traditional quantum field theory does not give expectation 
values but rather in-out matrix elements. On de Sitter background the
integration over ${x'}^{\mu}$ would be problematic for the in-out matrix 
element. To get a true expectation value one must employ the 
Schwinger-Keldysh formalism \cite{Schwinger:1960qe,Mahanthappa:1962ex,
Bakshi:1962dv,Bakshi:1963bn,Keldysh:1964ud,Chou:1984es,Calzetta:1986ey}. 
The rules are simple \cite{Ford:2004wc}:
\begin{itemize}
\item{Each endpoint of a propagator carries a $\pm$ polarity. The $++$
propagator is the same as the Feynman propagator while the $--$ propagator
is its complex conjugate. The $-+$ propagator is free vacuum expectation 
value of $\phi(x) \times \phi(x')$ while the $+-$ propagator is its
complex conjugate.}
\item{Interaction vertices, including counterterms, are either all $+$ 
or all $-$. The $+$ vertices are the same as for the in-out formalism 
while the $-$ vertices are complex conjugated.}
\item{External lines can be either $+$ or $-$. The $+$ lines correspond
to time-ordered fields while the $-$ lines correspond to anti-time-ordered
fields.}
\end{itemize}

On de Sitter background the four Schwinger-Keldysh propagators are all 
obtained from expression (\ref{propagator}) by making different
substitutions for the de Sitter length function $y(x;x')$. The $++$ case
is the same as (\ref{ydef}), and the $--$ case is its complex conjugate.
The $-+$ case is the complex conjugate of the $+-$ case, which is,
\begin{equation}
y_{\scriptscriptstyle +-}(x;x') \equiv a a' H^2 \Bigl[ \Vert \vec{x} \!-\! 
\vec{x}' \Vert^2 - (\eta \!-\! \eta' \!+\! i \epsilon)^2 \Bigr] \; . 
\label{y+-}
\end{equation}
Note that it agrees with $y_{\scriptscriptstyle ++}(x;x')$ for $\eta <
\eta'$, and it is the complex conjugate for $\eta > \eta'$. 

\subsection{Dimensionally Regulated Computation}

The analytic form of the diagram depicted in Figure~\ref{phisq} is,
\begin{eqnarray}
\lefteqn{ \int \!\! d^Dx' \sqrt{-g(x')} \Bigl[i 
\Delta^2_{\scriptscriptstyle ++}(x;x') - i \Delta^2_{\scriptscriptstyle 
+-}(x;x') \Bigr] \times \Bigl[-\tfrac{i}{2} \lambda i\Delta(x';x')
-i\delta \xi R(x')\Bigr] } \nonumber \\
& & \hspace{2.5cm} = -i k \lambda \!\! \int \!\! d^Dx' {a'}^D
\ln( \tfrac{\mu_1 a'}{H}) \Bigl[ i \Delta^2_{\scriptscriptstyle ++}(x;x') 
- i \Delta^2_{\scriptscriptstyle +-}(x;x') \Bigr] \; . \qquad 
\label{phisq1}
\end{eqnarray}
The next step is to substitute expression (\ref{propsq}) for the
squared propagators. Only the $++$ term contributes to the divergence,
which gives,
\begin{equation}
\Bigl({\rm Figure}~\ref{phisq}\Bigr)_{\rm div} = \tfrac{k \lambda 
\mu^{D-4}}{8 \pi^{\frac{D}2}} \tfrac{\Gamma(\frac{D}2 - 1)}{(D-3)(D-4)}
\tfrac{\ln(\frac{\mu_1 a}{H})}{a^{D-4}} \; . \label{phisqdiv} 
\end{equation}
This divergence requires composite operator renormalization because
$\phi^2(x)$ is a composite operator. The required counterterm is,
\begin{equation}
\Delta \phi^2 = \tfrac{\lambda \mu^{D-4}}{16 \pi^{\frac{D}2}}
\tfrac{\Gamma(\frac{D}2 -1)}{(D-3) (D-4)} \Bigl[-\phi^2(x) +
\tfrac{\Gamma(D-1) \Psi(D) \mu_1^{D-4} R(x)}{(4\pi)^{\frac{D}2} 
\Gamma(\frac{D}2) D (D-1)} \Bigr] \; . \label{composite} 
\end{equation}
The finite remainder is,
\begin{equation}
\Bigl({\rm Figure}~\ref{phisq}\Bigr)_{\rm rem} = 
\tfrac{\lambda H^2}{2^7 \pi^4} \Bigl\{-2 \ln(a) \ln(\tfrac{\mu_1 a}{H})
- \ln^2(\tfrac{\mu_1}{H}) - [2 \ln(2) - \tfrac52 + 2\gamma ] 
\ln(\tfrac{\mu_1}{H}) \Bigr\} \; . \label{remainder}
\end{equation}

The finite part consists of three terms, the first of which is,
\begin{eqnarray}
\lefteqn{\Bigl({\rm Figure}~\ref{phisq}\Bigr)_{\rm fin1} = 
\tfrac{i \lambda H^2 \partial^2}{2^9 \pi^6 a^2} \!\! \int \!\! d^4x' \,
{a'}^2 \ln(\tfrac{\mu_1 a'}{H}) \Bigl\{ \tfrac{\ln(\mu^2 \Delta x^2_{++})
}{\Delta x^2_{++}} - \tfrac{\ln(\mu^2 \Delta x^2_{+-})}{\Delta x^2_{+-}} 
\Bigr\} \; ,} \\
& & \hspace{-.5cm} = \tfrac{i \lambda H^2 \partial^4}{2^{12} \pi^6 a^2} 
\!\! \int \!\! d^4x' \, {a'}^2 \ln(\tfrac{\mu_1 a'}{H}) \Bigl\{ 
\ln^2(\mu^2 \Delta x^2_{\scriptscriptstyle ++}) \!-\! 2 \ln(\mu^2 
\Delta x^2_{\scriptscriptstyle ++}) - ({\scriptscriptstyle ++} 
\rightarrow {\scriptscriptstyle +-}) \Bigr\} \; , \qquad \\
& & \hspace{-.5cm} = -\tfrac{\lambda H^2 \partial^4}{2^{10} \pi^5 a^2} 
\!\! \int \!\! d^4x' \, {a'}^2 \ln(\tfrac{\mu_1 a'}{H}) \theta(\Delta 
\eta \!-\! \Delta r) \Bigl\{ \ln[\mu^2 (\Delta \eta^2 \!-\! \Delta r^2)] 
\!-\! 1\Bigr\} \; , \qquad \\
& & \hspace{-.5cm} = -\tfrac{\lambda H^2 \partial_0^4}{2^{8} \pi^4 a^2} 
\!\! \int_{\eta_i}^{\eta} \!\! d\eta' \, {a'}^2 \ln(\tfrac{\mu_1 a'}{H}) 
\Delta \eta^3 \Bigl\{\tfrac23 \ln(2 \mu \Delta \eta) \!-\! \tfrac{11}{9} 
\Bigr\} \; , \qquad \\
& & \hspace{-.5cm} = -\tfrac{\lambda H^2 \partial_0}{2^{6} \pi^4 a^2} 
\!\! \int_{\eta_i}^{\eta} \!\! d\eta' \, {a'}^2 \ln(\tfrac{\mu_1 a'}{H})
\times \ln(2 \mu \Delta \eta) \; , \qquad \\
& & \hspace{-.5cm} = -\tfrac{\lambda H^2}{2^{6} \pi^4} 
\tfrac{\partial}{\partial a} \!\! \int_{1}^{a} \!\! da' \, 
\ln(\tfrac{\mu_1 a'}{H}) \ln\Bigl[\tfrac{2 \mu}{H} (\tfrac1{a'} \!-\! 
\tfrac1{a}) \Bigr] \; , \qquad \\
& & \hspace{-.5cm} = \tfrac{\lambda H^2}{2^{6} \pi^4} \Bigl\{
\ln(\tfrac{H a}{2 \mu}) \ln(\tfrac{\mu_1 a}{H}) + \ln(\tfrac{\mu_1 a}{H})
- 1 + \tfrac{\pi^2}{6} + \sum_{n=2}^{\infty} \tfrac{1}{n a^n} \Bigl[
\ln(\tfrac{\mu_1}{H}) - \tfrac1{n}\Bigr] \Bigr\} \; . \qquad 
\label{fin1}
\end{eqnarray}
Note that the sub-leading logarithm --- $\lambda \ln^2(a)$ --- cancels
between (\ref{remainder}) and (\ref{fin1}). Both of these contributions 
derive from the first term in the expansion (\ref{propagator}) of the
propagator.

The 2nd and 3rd finite contributions are simplified if we make the 
definition,
\begin{equation}
\chi \equiv \exp[\gamma - 1] \! \times \! H \qquad \Longrightarrow 
\qquad \ln(H^2 \Delta x^2) + 2 \gamma - 2 = \ln(\chi^2 \Delta x^2) \; .
\label{chidef}
\end{equation}
The second of the finite contributions is,
\begin{eqnarray}
\lefteqn{\Bigl({\rm Figure}~\ref{phisq}\Bigr)_{\rm fin2} =
\tfrac{i \lambda H^4}{2^7 \pi^6 a} \!\! \int \!\! d^4x' \, {a'}^3
\ln(\tfrac{\mu_1 a'}{H}) \Bigl\{ \tfrac{\ln(\chi^2 \Delta x^2_{++})
}{\Delta x^2_{++}} - \tfrac{\ln(\chi^2 \Delta x^2_{+-})}{\Delta x^2_{+-}} 
\Bigr\} \; ,} \\
& & \hspace{-.5cm} = \tfrac{i \lambda H^4 \partial^2}{2^{10} \pi^6 a} 
\!\! \int \!\! d^4x' \, {a'}^3 \ln(\tfrac{\mu_1 a'}{H}) \Bigl\{ 
\ln^2(\chi^2 \Delta x^2_{\scriptscriptstyle ++}) \!-\! 2 \ln(\chi^2 
\Delta x^2_{\scriptscriptstyle ++}) - ({\scriptscriptstyle ++} 
\rightarrow {\scriptscriptstyle +-}) \Bigr\} \; , \qquad \\
& & \hspace{-.5cm} = \tfrac{\lambda H^2}{2^4 \pi^4 a} \!\! \int_{1}^{a} 
\!\! da' \, \ln(\tfrac{\mu_1 a'}{H}) (1 - \tfrac{a'}{a}) \Bigl\{
\ln\Bigl[ \tfrac{2 \chi}{H} (\tfrac1{a'} - \tfrac1{a})\Bigr] - 1
\Bigr\} \; , \qquad \\
& & \hspace{-.5cm} = -\tfrac{\lambda H^2}{2^5 \pi^4} \Bigl\{ 
\ln^2(\tfrac{\mu_1 a}{H}) + O\Bigl( \ln(\tfrac{\mu_1 a}{H}) \Bigr)
\Bigr\} \; . \label{fin2}
\end{eqnarray} 
The final finite contribution contains both leading and first 
sub-leading logarithms,
\begin{eqnarray}
\lefteqn{\Bigl({\rm Figure}~\ref{phisq}\Bigr)_{\rm fin3} = -
\tfrac{i \lambda H^6}{2^9 \pi^6} \!\! \int \!\! d^4x' \, {a'}^4
\ln(\tfrac{\mu_1 a'}{H}) \Bigl\{ \ln^2(\chi^2 \Delta x^2_{
\scriptscriptstyle ++}) - \ln^2(\chi^2 \Delta x^2_{\scriptscriptstyle +-}) 
\Bigr\} ,} \\
& & \hspace{1cm} = \tfrac{\lambda H^6}{2^7 \pi^5} \!\! \int \!\! d^4x' 
\, {a'}^4 \ln(\tfrac{\mu_1 a'}{H}) \theta(\Delta \eta - \Delta r) 
\ln[\chi^2 (\Delta \eta^2 - \Delta r^2)] \; , \qquad \\
& & \hspace{1cm} = \tfrac{\lambda H^6}{2^5 \pi^4} \!\! \int_{\eta_i}^{\eta}
\!\! d\eta' \, {a'}^4 \ln(\tfrac{\mu_1 a'}{H}) \Delta \eta^3 \Bigl\{
\tfrac23 \ln(2 \chi \Delta \eta) - \tfrac89\Bigr\} \; , \qquad \\
& & \hspace{1cm} = \tfrac{\lambda H^2}{2^5 \pi^4} \!\! \int_{1}^{a} \!\!
\tfrac{da'}{a'} (1 - \tfrac{a'}{a})^3 \ln(\tfrac{\mu_1 a'}{H}) \Bigl\{
\tfrac23 \ln\Bigl[ \tfrac{2\chi}{H} (\tfrac1{a'} - \tfrac1{a})\Bigr] 
- \tfrac89 \Bigr\} \; , \qquad \\
& & \hspace{1cm} = \tfrac{\lambda H^2}{2^5 \pi^4} \Bigl\{-\tfrac29 
\ln^3(\tfrac{\mu_1 a}{H}) \!+\! \tfrac13 \Bigl[ \ln(\tfrac{2 \mu_1 \chi}{H^2})
\!+\! \tfrac{7}{3}\Bigr] \ln^2(\tfrac{\mu_1 a}{H}) \!+\! O\Bigl( 
\ln(\tfrac{\mu_1 a}{H}) \Bigr) \Bigr\} \; . \qquad \label{fin3}
\end{eqnarray}
It follows that the order $\lambda$ contribution to the expectation 
value of $\phi^2(x)$ is,
\begin{eqnarray}
\lefteqn{\tfrac{\lambda H^2}{2^5 \pi^4} \Bigl\{-\tfrac29 
\ln^3(\tfrac{\mu_1 a}{H}) \!+\! \tfrac13 \Bigl[ \ln(\tfrac{2 \mu_1 
\chi}{H^2}) \!-\! \tfrac{2}{3} \Bigr] \ln^2(\tfrac{\mu_1 a}{H}) \!+\! 
O\Bigl( \ln(\tfrac{\mu_1 a}{H}) \Bigr) \Bigr\} } \nonumber \\
& & \hspace{1.5cm} = -\tfrac{\lambda H^2}{2^4 3^2 \pi^4} \ln^3(a) -
\tfrac{\lambda H^2}{2^5 3^1 \pi^4} \Bigl[\ln(\tfrac{\mu_1}{2 H})
\!+\! \tfrac53 \!-\! \gamma \Bigr] \ln^2(a) + O\Bigl( \ln(a)\Bigr) 
. \qquad \label{VEVphisq}
\end{eqnarray}
Of course the order $\lambda \ln^3(a)$ term agrees with the stochastic
prediction (\ref{VEVsquare}). The coefficient of the $\lambda \ln^2(a)$ 
part of the stochastic prediction (\ref{VEVsquare}) contains the same
crucial factor of $\ln(\frac{\mu_1}{2 H})$ as the exact computation 
(\ref{VEVphisq}), but it lacks the $\frac53 - \gamma$. A likely resolution
to this disagreement is to make a small adjustment in the lower limit of
the stochastic mode sum (\ref{stochmodesum}),
\begin{equation}
H \longrightarrow \exp\Bigl[\tfrac56 - \tfrac{\gamma}{2}\Bigr] \times H
\simeq 1.724 \times H \; . \label{shift}
\end{equation}  

\section{Conclusions}

The first sub-leading logarithm is crucially important in the theory
of $\Lambda$-driven inflation because it describes primordial 
perturbations around the homogeneous, leading logarithm background.
In this paper we have studied the first sub-leading logarithm in
the simpler context of scalar potential models. Because reaching 
leading logarithm order requires every pair of free fields to produce
a factor of $\ln[a(t)]$, the first sub-leading logarithm should be 
attained by ``wasting'' a single pair of free fields. We argued that 
this is accomplished by a certain part of the 1-loop correction to 
the effective potential. In section 2 we derived the two counterterms
(\ref{deltaxi}) and (\ref{deltalambda}) needed to renormalize the 
1-loop effective potential. The derivative of the effective potential
(\ref{complete}) was duly computed in section 3, and then used in the
stochastic formalism to predict the order $\lambda$ correction to the
expectation value of $\phi^2$ (\ref{VEVsquare}). This prediction was
checked in section 4 with a dimensionally regulated and fully 
renormalized computation (\ref{VEVphisq}). The two results agree up 
to a small shift (\ref{shift}) in the lower limit of the stochastic
mode sum (\ref{stochmodesum}).

Several points deserve comment. The first is the role of ultraviolet
divergences. The leading logarithms of scalar potential models are
completely ultraviolet finite, but that this no longer true for the
first sub-leading logarithms. One can see this from the dependence
of the results (\ref{VEVsquare}) and (\ref{VEVphisq}) on the 
dimensional regularization scale $\mu_1$ of the conformal counterterm
(\ref{deltaxi}). However, no other counterterms affect the first
sub-leading correction (\ref{1stSub}). In particular, the vertex
counterterm (\ref{deltalambda}) does not matter, nor is there any
dependence on the composite operator counterterm (\ref{composite}) 
needed to renormalize the expectation value of $\phi^2(x)$. 

A second comment concerns the fact that only a small portion
(\ref{1stSub}) of the derivative of the 1-loop effective potential 
(\ref{complete}) contributes to the first sub-leading logarithm. The
full result (\ref{complete}) actually contains leading logarithm
terms (\ref{Leading}), as well as even more sub-dominant logarithms.
The reason for this is that the effective potential can be viewed as
the sum of 1PI vacuum graphs such as Figure~\ref{Vsum}. The $N$-point
contribution consists of $N$ factors of $\lambda \phi^2$ with $N$
propagators. In the stochastic formalism this $\lambda^N$ contribution 
will receive $N$ factors of $\ln[a(t)]$ from the fields. Each of the 
$N$ propagators can also produce a factor of $\ln[a(t)]$, or not. The
result is at leading logarithm when all $N$ propagators contribute a
factor of $\ln[a(t)]$, it is at first sub-leading order when only 
$N-1$ contribute, and so on.

As the preceding paragraph has explained, the derivative of the 
effective potential (\ref{complete}) is highly time dependent. This time
dependence is crucial to allowing higher powers of $\lambda \phi^2$ 
to contribute at leading and first sub-leading logarithm order.
This phenomenon is crucial to understanding an issue that occurs in
theories with derivative interactions of scalars or gravitons. In
these models one can get leading logarithm factors of $\ln[a(t)]$
from the infrared sector, just as for scalar potential models. 
However, there is also another type of leading logarithm arising
from incomplete cancellation between primitive divergences and
counterterms \cite{Miao:2021gic}.\footnote{The current work actually
shows the same incomplete cancellation between the primitive
divergence (\ref{phisqdiv}) and the composite operator renormalization
(\ref{composite}). However, this is not a leading logarithm effect in
scalar potential models.} Whereas the first sort of logarithms can be
described stochastically by a Langevin equation, the second sort 
requires a variant of the renormalization group. Many examples of the
second sort have been encountered in nonlinear sigma models 
\cite{Miao:2021gic} and in gravity \cite{Woodard:2025smz}. The issue
is mixing between the two effects, when the leading logarithms at 
$\ell$-loop come partly from one source and partly from the other. 
We suspect that this can be handled by using the Callan-Symanzik
equation to renormalization group improve the effective potential,
and then employing the improved potential in the stochastic Langevin
equation. Note that the improvement terms will be time-dependent in
roughly the same way that (\ref{complete}) is.

\centerline{\bf Acknowledgements}

This work was partially supported by Taiwan NSTC grants 
113-2112-M-006-013 and 114-2112-M-006-020, by NSF grant 
PHY-2207514 and by the Institute for Fundamental Theory 
at the University of Florida.

\end{document}